\documentstyle[12pt]{article}

\begin{document}
{\sf \begin{center} \noindent {\Large\bf Lyapunov spectra in fast dynamo Ricci flows of negative sectional curvature}\\[2mm]

by \\[0.1cm]

{\sl  L.C. Garcia de Andrade}\\Departamento de F\'{\i}sica
Te\'orica -- IF -- Universidade do Estado do Rio de Janeiro-UERJ\\[-3mm]
Rua S\~ao Francisco Xavier, 524\\[-3mm]
Cep 20550-003, Maracan\~a, Rio de Janeiro, RJ, Brasil\\[-3mm]
Electronic mail address: garcia@dft.if.uerj.br\\[-3mm]
\vspace{1cm} {\bf Abstract}
\end{center}
\paragraph*{}
Previously Chicone, Latushkin and Montgomery-Smith [\textbf{Comm.
Math. Phys. \textbf{173},(1995)}] have investigated the spectrum of
the dynamo operator for an ideally conducting fluid. More recently,
Tang and Boozer [{\textbf{Phys. Plasmas (2000)}}], have investigated
the anisotropies in magnetic field dynamo evolution, from
finite-time, Lyapunov exponents, giving rise to a Riemann metric
tensor, in the Alfven twist in magnetic flux tubes (MFTs). In this
paper one investigate the role of Perelman Ricci flows constraints
in twisted magnetic flux tubes, where the Lyapunov eigenvalue
spectra for the Ricci tensor associated with the Ricci flow equation
in MFTs leads to a finite-time Lyapunov exponential stretching along
the toroidal direction of the tube and a contraction along the
radial direction of the tube. It is shown that in the case of MFTs,
the sectional Ricci curvature of the flow, is negative as happens in
geodesic flows of Anosov type. Ricci flows constraints in MFTs
substitute the Thiffeault and Boozer [\textbf{Chaos}(2001)] have
vanishing of Riemann curvature constraint on the Lyapunov
exponential stretching of chaotic flows. Gauss curvature of the
twisted MFT is also computed and the contraints on a negative Gauss
curvature are obtained.

{\bf PACS numbers:\hfill\parbox[t]{13.5cm}{02.40.Hw:differential
geometries.91.25.Cw-dynamo theories.}}}
\newline
\newpage
 \section{Introduction}
 After the Perelman's seminal paper \cite{1}, on an explanation of the Poincare conjecture, many applications of the Ricci flow equations to several
 areas of physics have appeared in the literature \cite{2,3}. More recently
 Cao \cite{3} has computed the first eigenvalues, of "heat flow"
operator
 $-{\Delta}+\frac{R}{2}$, where R is the Ricci curvature scalar and ${\Delta}:={\nabla}^{2}$ is the Laplacian 3D operator, under Ricci flow. He found that the eigenvalues were
 non-decreasing as happens in general with Lyapunov exponents \cite{4}. In this paper one investigates a similar problem in the context of the time evolution
 operator under Ricci flow in the backyard of the twisted MFT. To be able to determine the Lyapunov eigenvalue spectra in twisted MFTs, which
 is a fundamental problem which helps one to determine the stability of the flows inside flux tubes such
 as tokamaks or stellarators in plasma physics \cite{5} or even in
 the context of solar or other stellar plasma loop \cite{6}, one computes the eigenvalue spectra of the
 Ricci tensor under the Ricci flow. This is an important problem for
 plasma theorists and experimentalists, shall be examined here in the
 framework of Riemannian geometry
 \cite{7}. Recently, Thiffeault, Tang and Boozer, investigated Riemannian constraints on Lyapunov exponents \cite{8}, based on the relation between the
 Riemann metric and the finite-time Lyapunov exponential stretching,
 so fundamental for dynamo action. Note that, here, another
 sort of constraint is investigated. Instead of the vanishing of the Riemann curvature tensor, called by mathematicians, Riemann-flat space or condition, one use
 the Lyapunov spectra under Ricci flow. Chaotic flows inside the twisted
 MFTs are investigated. Anosov diffeomorpism \cite{9}, is an important mathematical tool from the theory of
 dynamical systems, that has often been used, in connection with
 the investigation of dynamo flows and maps \cite{10} such as the Arnold's Cat Map \cite{10} on the torus, useful in mixing \cite{11} problems in the
 physics of fluids \cite{11}. One of the main properties of
 the Anosov maps is that they yield Lyapunov exponential of the
 chaotic exponential stretching, which are constant everywhere \cite{12}. This paper is organized as follow: In section 2 the dynamo maps under Ricci
 flows in MFTs are investigated with the aid of Lyapunov spectra. In section 3, the thin tube perturbations are computed in the negative Ricci sectional
 curvature assumption. Section 4 addresses discussions and conclusions.
 \newpage
\section{Ricci fast magnetic dynamo flows in MFTs}
\vspace{1cm} Let us start this section, by defining the Ricci flow
as: \newline \textbf{Definition 2.1}: \newline Let us consider a
smooth manifold which Ricci tensor $\textbf{Ric}$ obeys the
following equation: \begin{equation}
\frac{{\partial}\textbf{g}}{{\partial}t}=2\textbf{Ric} \label{1}
\end{equation}
Here $\textbf{g}$ is the Riemann metric over the manifold $\cal{M}$
where in $\textbf{g}(t)$, $t{\in} [a,b]$. By chosing a local chart
$\cal{U}$ on this manifold, the Ricci flow equation may be written
as
\begin{equation}
\frac{{\partial}{g}_{ij}}{{\partial}t}=-2{R}_{ij} \label{2}
\end{equation}
with this equation in hand let us now compute the eigenvalue spectra
of the Ricci tensor as
\begin{equation}
{R}_{ij}{\chi}^{j}={\lambda}{\chi}_{i} \label{3}
\end{equation}
where here $(i,j=1,2,3)$. Substitution of the Ricci flow equation
(\ref{2}) into the eigenvalue equation (\ref{3}) one obtains an
eigenvalue equation for the metric itself, as
\begin{equation}
\frac{{\partial}{g}_{ij}}{{\partial}t}{\chi}^{j}=-2{\lambda}{g}_{ij}{\chi}^{j}
\label{4}
\end{equation}
and since the eigenvector ${\chi}^{k}$ is in principle arbitrary,
one can reduce this equation to
\begin{equation}
\frac{{\partial}{g}_{ij}}{{\partial}t}=-2{\lambda}{g}_{ij} \label{5}
\end{equation}
which reduces to the solution
\begin{equation}
{g}_{ij}=exp[-2{\lambda}_{i}t]{\delta}_{ij} \label{6}
\end{equation}
one notes that the, ${\delta}_{ij}$ is the Kroenecker delta diagonal
unity matrix. No Einstein sum convention is being used here. By
considering the Tang-Boozer relation between the metric $g_{ij}$
components and the Lyapunov exponents
\begin{equation}
g_{ij}={\Lambda}_{1}\textbf{e}_{1}\textbf{e}_{1}+{\Lambda}_{2}\textbf{e}_{2}\textbf{e}_{2}+{\Lambda}_{3}\textbf{e}_{3}\textbf{e}_{3}
\label{7}
\end{equation}
\newpage
the following lemma, can be proved:
\newline
\textbf{lemma 1}: \newline If ${\lambda}_{i}$ is the eigenvalue
spectra of the Ricci tensor $\textbf{Ric}$ under Ricci flow
equation, the Lyapunov spectra is given by the following relations:
\begin{equation}
{\lambda}_{i}=-{{\gamma}_{i}}\le{0} \label{8}
\end{equation}
where ${\lambda}_{i}$ are the finite-time Lyapunov numbers. The
infinite or true Lyapunov number is
\begin{equation}
{{\lambda}_{i}}^{\infty}=\lim_{{t\rightarrow\infty}}(\frac{ln{\Lambda}_{i}}{2t})
\label{9}
\end{equation}
Here one has used the finite-time Lyapunov exponent given
by
\begin{equation}
{{\lambda}}_{i}=(\frac{ln{\Lambda}_{i}}{2t}) \label{10}
\end{equation}
Note that one of the interesting features of the Ricci flow method
is that one may find the eigenvalue Lyapunov spectra without
computing the Ricci tensor, of the flux tube, for example. Actually
the twisted flux tube Riemannian line element \cite{13}
\begin{equation}
dl^{2}= dr^{2}+r^{2}d{{\theta}_{R}}^{2}+K^{2}(r,s)ds^{2} \label{11}
\end{equation}
can now be used, to compute the Lyapunov exponential stretching of
the flow as in Friedlander and Vishik \cite{14} "dynamo flow" with
the Ricci flow technique above.  Here twist transformation angle is
given by
\begin{equation}
{\theta}(s):={\theta}_{R}-\int{{\tau}(s)ds} \label{12}
\end{equation}
One another advantage of the method used here is that this allows us
to compute the finite-time Lyapunov exponential, without the need of
recurring to the non-Anosov maps \cite{15}. The Lyapunov exponents
here are naturally non-Anosov since the exponents are
non-homogeneous. Here $K(r,s):=(1-r{\kappa}(s,t)cos{\theta})$, is
the stretching in the metric. Let us now compute the eigenvalue
spectra for the MFTs. Note that the eigenvalue problem, can be
solved by the $3D$ matrix
\begin{equation}
\textbf{M}_{\textbf{3D}}=\pmatrix{2{\lambda}g_{11}&0&0\cr0&{\partial}_{t}g_{22}+{\lambda}g_{22}&0\cr0&0&{\partial}_{t}g_{33}+{\lambda}g_{33}\cr}\qquad
\label{13}
\end{equation}
The eigenvalue equation
\begin{equation}
Det[\textbf{M}_{\textbf{3D}}]=0 \label{14}
\end{equation}
the following eigenvalue Lyapunov spectra for thick tubes, where
$K\approx{-{\kappa}_{0}rcos{\theta}(s)}$ are
\begin{equation}
{\lambda}_{1}=0, {\lambda}_{2}=
2\frac{v_{r}(r)}{r},{\lambda}_{3}=\frac{1}{2}{\lambda}_{2}+{\omega}_{1}tg{\theta}(s)
\label{15}
\end{equation}
Therefore, since for the existence of dynamo action, at least two of
the Lyapunov exponents have to have opposite signs \cite{15} in
order to obey the stretching $({\lambda}_{3}>0)$ and contracting
$({\lambda}_{2}<0)$, in order that the radial flow ${v}_{r}$ be
negative, ${\lambda}_{3}>0$. Therefore from the above expressions
the following constraint is obtained:
\begin{equation}
|{\omega}_{1}tg{\theta}(s)|\ge{|{\lambda}_{2}|}=|\frac{v_{r}}{r}|
\label{16}
\end{equation}
Thus the $\frac{{v}_{r}}{r}<0$ yields a compression on the flux tube
which induces the in the tube a stretch along the toroidal
direction-s, by the stretch-twist and fold dynamo generation method
of Vainshtein and Zeldovich \cite{16}. In the next section one shall
compute the relation between twist or vorticity, and the when the
sectional curvature of the thin tube is negative. Actually this
leads us to the following time dependence of the magnetic field
components, as
\begin{equation}
{B}_{\theta}\approx{e^{2{\lambda}_{\theta}t}}=e^{(\frac{v_{r}}{r})t}
\label{17}
\end{equation}
\begin{equation}
B_{s}\approx{e^{{\lambda}_{s}t}}=e^{(\frac{v_{r}}{r}+{\omega}_{1}v_{r}tan{\theta})t}
\label{18}
\end{equation}
where ${\omega}_{1}$ is a constant vorticity inside the dynamo flux
tube.

\section{Magnetic dynamo flows of negative sectional curvature}
In this section the Ricci sectional curvature \cite{17} in the case
of a radial perturbation of a thin twisted MFT. This is justified
since, as one has seen in the last section, the radial flow is
fundamental for the existence of non-vanishing Lyapunov exponential
stretching, which in turn are fundamental for the existence of
dynamo action. On the other hand, following work by D. Anosov
\cite{9}, Chicone and Latushkin \cite{18} have previously shown that
geodesic flows, which possesses negative Riemannian curvature has a
fast dynamo action. Let X and Y be vectors laying in tangent
manifolds $\cal{TM}$ to a Riemannian manifold
$\cal{M}\subset{\cal{N}}$ where $\cal{N}$ is an Euclidean three
dimensional space. The Ricci sectional curvature is given by
\begin{equation}
K(X,Y):=\frac{<R(X,Y)Y,X>}{S(X,Y)} \label{19}
\end{equation}
where $R(X,Y)Z$ is the Riemann curvature given by
\begin{equation}
R(X,Y)Z={\nabla}_{X}{\nabla}_{Y}Z-
{\nabla}_{Y}{\nabla}_{X}Z-{\nabla}_{[X,Y]}Z \label{20}
\end{equation}
where
\begin{equation}
S(X,Y):=||{X}||^{2}||Y||^{2}-<X,Y>^{2} \label{21}
\end{equation}
As usual ${\nabla}_{X}Y$ is the Riemannian covariant derivative
given by
\begin{equation}
{\nabla}_{X}Y=({X}.{\nabla})Y \label{22}
\end{equation}
Expression $[X,Y]$ is the commutator, where which on the vector
frame $\textbf{e}_{l}$, $(l=1,2,3)$ in $\textbf{R}^{3}$, as
\begin{equation}
{X}={X}_{k}\textbf{e}_{k}\label{23}
\end{equation}
or its dual basis
\begin{equation}
{X}={X}^{k}{\partial}_{k}\label{24}
\end{equation}
where Einstein summation convention is used. Thus the commutator is
written as
\begin{equation}
[X,Y]={[X,Y]}^{k}{\partial}_{k} \label{25}
\end{equation}
Thus the Riemann curvature tensor becomes
\begin{equation}
R(X,Y)Z=[{R^{l}}_{jkp}Z^{j}X^{k}Y^{p}]{\partial}_{l} \label{26}
\end{equation}
By considering the thin tube approximation $K\approx{1}$ where the
gradient is given by
\begin{equation}
{\nabla}=[{\partial}_{r},r^{-1}{\partial}_{{\theta}_{R}},{\partial}_{s}]
\label{27}
\end{equation}
Let the radial perturbation in the manifold of flux tube be given by
\begin{equation}
X={v^{l}}_{r}\textbf{e}_{r} \label{28}
\end{equation}
where it was considered that the background radial flow vanishes.
The vector Y is given by
\begin{equation}
Y={v}_{\theta}\textbf{e}_{\theta}+v_{s}\textbf{t} \label{29}
\end{equation}
where the Frenet vector $\textbf{t}$ is tangent to the magnetic tube
axis, and the perturbation is chosen along the radial direction,
because in general flux tubes initially possesses a strongly
confined magnetic field along the tube. In the way of computing the
$\textbf{Ric}$, the covariant derivative is
\begin{equation}
{\nabla}_{X}Y={v^{1}}_{r}{\partial}_{r}[v_{\theta}\textbf{e}_{\theta}+v_{s}\textbf{t}]
\label{30}
\end{equation}
which, under the approximations
\begin{equation}
{v^{1}}_{r}{\partial}_{r}v_{\theta}\approx{0} \label{31}
\end{equation}
\begin{equation}
{v^{1}}_{r}{\partial}_{r}v_{s}\approx{0} \label{32}
\end{equation}
The MFT relation
\begin{equation}
{\nabla}_{Y}{\nabla}_{X}Y\approx{0} \label{33}
\end{equation}
Since the term
\begin{equation}
{\nabla}_{[X,Y]}Y=O(v^{3}) \label{34}
\end{equation}
and one is assuming that the velocities involved in the plasma
dynamo flow are nor small neither turbulent, this term can be
dropped and the $\textbf{Ric}$ can be expressed as
\begin{equation}
R(X,Y)Y\approx{{\nabla}_{X}{\nabla}_{Y}Y} \label{35}
\end{equation}
This yields
\begin{equation}
R(X,Y)Y\approx{[v_{s}-{{\tau}(s)}^{-1}][v_{\theta}{\kappa}{\tau}sin{\theta}\textbf{e}_{\theta}-{\tau}v_{\theta}sin{\theta}\textbf{t}+v_{s}{\kappa}
\textbf{n}]} \label{36}
\end{equation}
Therefore after some algebraic manipulation one obtains the
sectional curvature as
\begin{equation}
K(X,Y):=\frac{{\kappa}(s)cos{\theta}}{{v^{1}}_{r}v_{s}} \label{37}
\end{equation}
Where one has considered the approximation that due to dynamo action
$v_{s}>>v_{\theta}$ which is usual, for example, in solar physics.
In the above computations one has considered the flow as
incompressible, ${\nabla}.\textbf{v}=0$, or obeying the equation
\begin{equation}
{\partial}_{s}v_{\theta}:={\kappa}{\tau}rsin{\theta}v_{\theta}
\label{38}
\end{equation}
The Gauss curvature is given by
\begin{equation}
K_{G}:=\frac{R_{1212}}{g} \label{39}
\end{equation}
Here $g:=det{g_{ij}}$ and $R_{1212}$ is the Riemann curvature of the
twisted MFT surface, given by the line element
\begin{equation}
dl^{2}= {r_{0}}^{2}d{{\theta}_{R}}^{2}+K^{2}(s)ds^{2} \label{40}
\end{equation}
This metric form has been obtained from expression (\ref{11}) by
simply consider the Riemann curved twisted flux tube surface of
constant cross-section of radius, $r=r_{0}=constant$. The Riemann
curvature $R_{1212}$ of this Riemannian line element is
\begin{equation}
R_{1212}=-{\kappa}(s)K(s)cos{{\theta}} \label{41}
\end{equation}
Here one have considered that the stretching metric coefficient is
$K(s)=(1-r_{0}{\kappa}(s)cos{\theta})$. Thus since $g={r_{0}}^{2}$,
substitution of this values into the $K_{G}$ Gauss curvature above
yields
\begin{equation}
K_{G}= -\frac{{\kappa}(s)K(s)cos{\theta}}{r_{0}} \label{42}
\end{equation}
This is a general form of the Gaussian curvature, which can now be
applied to the case of thin constant Frenet curvature MFTs, by
making $K\approx{1}$ and ${\kappa}={\kappa}_{0}=constant$, which
yields
\begin{equation}
K_{G}= -\frac{{\kappa}_{0}cos{\theta}}{r_{0}} \label{43}
\end{equation}
Note from this expression that, for it be negative, the
$cos{\theta}>0$ and the Frenet curvature of the magnetic flux tube
axis, positive, or they have to have the same sign at all. Just for
comparison one mention here the fast kinematic dynamo eigenvalue
spectrum, in diffusive media obtained by Chicone and Latushikin
\cite{18}, which is
\begin{equation}
{\lambda}_{\epsilon}=
\frac{1}{2}[-{\epsilon}(1+{\kappa}^{2})+\sqrt{{\epsilon}^{2}(1-{\kappa}^{2})^{2}-4{\kappa})}]
\label{44}
\end{equation}
where ${\epsilon}$ is the resistive plasma coefficient of the
diffusive fast kinematic dynamo represents the growth rate of the
magnetized plasma dynamo \cite{19}, and ${\lambda}_{\epsilon}$ In
the case of diffusive-free $({\epsilon}=0)$ ideal plasma one obtains
\begin{equation}
{\lambda}_{0}= i[\sqrt{{\kappa}}] \label{45}
\end{equation}
which is a pure magnetic unity. This is similar to the
${\lambda}_{3}$ one has obtained from the eigenvalue matrix above
that the ${\lambda}_{3}=2{\kappa}rcos{\theta}$ it is also
proportional to the Frenet curvature.
\section{Conclusions}
Let us state here for discussion the Chicone-Latushkin fast dynamo
in geodesic flow theorem \cite{18} as:
\newline
\textbf{Theorem}: If $\textbf{v}$ is the vector field that generates
the geodesic flow for a closed two dimensional Riemannian manifold
$\cal{M}$ . then $\textbf{v}$ is a steady solution of Euler´s
equation on $\cal{M}$. In addition if $\cal{M}$ has constant
negative curvature $\kappa$, then for each magnetic Reynolds number
$Re_{m}>\sqrt{-{\kappa}}$, the corresponding dynamo operator has a
positive eigenvalue given by ${\lambda}_{\epsilon}$ above.
\newline
Therefore one may say that in this paper we gu=ive an example of the
validity of Chicone-Latushkin theorem on geodesic dynamo flows in
the case of flux tubes where a containing R icci plasma flow. Also
in this paper the importance of investigation the Ricci flows as a
constraint to dynamo flows inside twisted MFTs is stressed, given
examples of the dynamo action existence in negative sectional
curvature. This can be basically done with the help of the Lyapunov
spectraof the Ricci fast dynamo flows, which are fundamental for the
exponenial stretching which are in turn so important for dynamo
action. An instability of the Lyapunov exponential stretching
influence on the instability of the Euler equations has been
discussed by Friedlander and Vishik \cite{14}. Physical implications
to the ideas discussed here to the dynamo magnetic flux tube
discussed and developed by Schuessler \cite{20} may appear
elsewhere.\section{Acknowledgements}:
\newline I am very much in debt to Jean Luc Thiffeault for calling
to my attention, many aspect of non-Anosov maps and stretching in
chaotic flows. I also am deeply indebt to G. Paternain , R. Ricca
and Dmitry Sokoloff for helpful discussions on the subject of this
paper. Financial supports from Universidade do Estado do Rio de
Janeiro (UERJ) and CNPq (Brazilian Ministry of Science and
Technology) are highly appreciated.


\begin{thebibliography}{24}
\bibitem{1} G. Perelman, The entropy formula for the Ricci flow and its geometric applications, Los Alamoa Arxives: math/0211599.
\end{thebibliography}
\end{document}